# Fabrication, Dynamics, and Electrical Properties of Insulated SPM Probes for Electrical and Electromechanical Imaging in Liquids

B.J. Rodriguez,[1,2] S. Jesse,[1] K. Seal,[2] A.P. Baddorf,[2] S.V. Kalinin,[*,1,2] and P. Rack[†,2]

[1]Materials Science and Technology Division and the [2]Center for Nanophase Materials Sciences, Oak Ridge National Laboratory, Oak Ridge, TN 37831

Insulated cantilever probes with a high aspect ratio conducting apex have been fabricated and their dynamic and electrical properties analyzed. The cantilevers were coated with silicon dioxide and a via was fabricated through the oxide at the tip apex and backfilled with tungsten to create an insulated probe with a conducting tip. The stiffness and Q-factor of the cantilevers increased after the modifications and their resonances shifted to higher frequencies. The coupling strength between the cantilever and the coating are determined. The applications to conductive and electromechanical imaging of ferroelectric domains are illustrated, and a probe apex repair process is demonstrated.

PACS: 07.79.Lh, 81.16.–c, 77.80.Dj

[*] Corresponding author, sergei2@ornl.gov
[†] Corresponding author, prack@utk.edu

The ubiquity of electromechanical phenomena in biological and biomolecular systems ranging from voltage controlled muscular contractions,[1] to cell electromotility,[2] ion channels, and electromotor proteins[3] necessitate probing these phenomena on the tissue, cellular, sub-cellular, and molecular levels. Scanning probe microscopy (SPM) studies of biological systems necessitate the use of a liquid environment to maintain a native environment for biomolecules and cells, to control tip-surface forces, and to eliminate capillary interactions. For electromechanical measurements, an additional degree of complexity is introduced by the conductivity of the liquid medium, resulting in stray currents and electrochemical reactions, thus precluding precise control of dc and ac probe potentials.

The use of high (0.1 – 25 MHz) imaging frequencies combined with direct or frequency-mixing detection allows one to avoid electrochemical processes for piezoelectric[4] and dielectroforetic force[5] imaging. However, the large response times (~ 0.1 -10 ms) of biological systems, combined with the requirement for precise control of local electrochemical potentials, require dc potential localization in solution within the probed volume. By using model ferroelectric systems and conventional metal-coated probes, it was shown that the electric field can be localized only in weakly conductive solvents such as isopropanol,[6] whereas even distilled water results in a delocalized dc field. These considerations necessitate the development of insulated SPM probes,[7,8,9] which effectively combine probe microscopy and patch-clamp techniques.[1]

The application of insulating coated probes to electromechanical and electrical probing in liquid environments requires (a) good dynamic properties and reflectivity of the lever, (b) good insulation except for the apex, (c) high apex conductivity, and (d) an apex geometry consistent with high resolution. In this Letter, we describe a process for fabricating shielded probes, the effect of the coating on the dynamic and electrical properties, and operation in conductive and electromechanical imaging modes in ambient and liquid environments.

Commercial doped-Si atomic force microscope (AFM) cantilevers (tip 1 – RFESP (Veeco) and tip 2 – AC240TS (Olympus)) were initially coated on both sides with ~ 500 nm silicon oxide by a plasma enhanced chemical vapor deposition process. The specific processing parameters were: 350 $^{o}$C, 85 sccm Ar(95%)-SiH$_4$(5%), and 157 sccm N$_2$O, and a total processing pressure of 1 Torr, 20 Watts radio-frequency power, for 7 minutes in an Oxford plasma enhanced chemical vapor deposition system. To allow for electrical contact, Kapton tape was placed over ~ 2 mm of the backside of the chip during oxide growth. Subsequent AFM tip processing was performed in an FEI Nova 600 dual beam scanning electron microscope/focused ion beam (FIB). To etch a via through the insulating oxide to the underlying silicon tip, a 100 nm diameter circle pattern was FIB milled into the AFM tip using a 30 keV, 30 pA gallium ion beam for 25 seconds. The resulting diameters were ~ 315 nm for tip 1 and ~ 240 nm for tip 2. Scanning electron microscope (SEM) images of tip 1 and tip 2 after the oxide deposition and after the via FIB are shown in Figs. 1(a,b) and 1(d,e), respectively. Some peripheral etch damage is observed in tip 1 which occurred during acquisition of an ion beam image. Subsequent to the via etch, a "tungsten" contact was deposited in the via using electron beam induced deposition (EBID).[10] A 10 keV (20 pA) electron beam in spot mode was used and the deposition precursor was W(CO)$_6$. Tip 1 was deposited for 35 s and tip 2 was deposited for 20 s. Typical EBID structures deposited from the W(CO)$_6$ precursor are ~ 55 at.% W, 30%C and 15% O.[11] SEM images of tip 1 and tip 2 after the EBID fill are shown in Figs. 1(c,f), respectively.

Dynamic characteristics of the cantilever, tip-surface contact resistance, and piezoresponse force microscopy (PFM) imaging in ambient and liquid environments were measured using an Asylum Research MFP-3D AFM with an additional function generator and lock-in amplifier (DS 345 and SRS 830, Stanford Research Systems). The scanner head was modified to allow direct access to the tip deflection signals. Before and after the processing, the photodetector sum signal, force-distance curves and thermal tunes were measured and recorded. The contact resistance and current-voltage (I-V) curves were measured on a clean Au(111) surface. The PFM imaging[12] was performed using periodically poled lithium niobate (PPLN) as a model system.

The tips were calibrated before and after processing using established methods to obtain total reflected light intensity (Sum) signal, spring constants, Q-factor, and inverse optical lever sensitivity (InvOLS) [Table 1].[13,14,15] The first thermal resonance before and after processing for tip 1 is shown in Figs. 2(a), and demonstrate a shift in the cantilever's first free resonance to higher frequencies. Note that $SiO_2$ deposition does not affect the reflectivity of the lever. The spring constant, $k$, and resonance frequency of the cantilever are increased. This behavior can easily be understood given that a rectangular cantilever spring constant is related to the geometric parameters of the cantilever as $k = 3EI/L^3 = Ewh^3/4L^3$. The resonant frequencies are $\omega_n^2 = k\beta_n^4/3m$, where $m$ is the cantilever mass and $\beta_n$ are the roots of the characteristic equation. For a free cantilever, the first three resonances occur for $\beta_n$ = 1.875, 4.694, 7.855 (compared to $\beta_n$ = 3.927, 7.067, 10.21 in contact).[16,17,18]

From the data in Table 1, the increase in effective mass for tip 1 is 22.9% and for tip 2 is 55%. From the known densities of Si and $SiO_2$ (2.33 and 2.2 g/cm$^3$, respectively), the increases in thickness are 24.2% and 58.2%, respectively. For the compound beam formed by a central beam of width $w_1$ and thickness $h_1$ surrounded by an external shell of width $w_2$ and thickness $h_2$, the effective stiffness is

$$12EI = E_1 h_1^3 w_1 + \alpha E_2 \left(h_2^3 w_2 - h_1^3 w_1\right), \quad (1)$$

where $\alpha$ is the constant ($0 \leq \alpha \leq 1$) describing the binding between the internal and outer beams. Ignoring the changes in the effective widths, Eq. (1) simplifies as $k^*/k = 1 + \alpha \{(h_2/h_1)^3 - 1\} E_2/E_1$. Using $E_2 = 70$ GPa for $SiO_2$ and $E_1 = 150$ GPa for Si, the coefficient is determined as $\alpha = 0.85$ for tip 1 and $\alpha = 0.74$ for tip 2.

The changes in the quality factor are analyzed using the Sader formulae, $k = 0.1906 w L^2 \rho_f Q_f \Gamma_i(\mathrm{Re}) \omega_f^2$, where $\rho_f$ is the density of the medium and $\Gamma_i(\mathrm{Re})$ is the hydrodynamic function. The Reynolds number is $\mathrm{Re} = \rho \omega w^2 / 4\eta$, where $\eta = 1.86 \cdot 10^{-5}$ kg/m s is the specific viscosity of air. Hence, the Q-factor after deposition, $Q_f^*$, is

$$Q_f^* = Q_f \frac{k^*}{k} \frac{w \Gamma_i(\mathrm{Re})}{w^* \Gamma_i(\mathrm{Re}^*)} \left(\frac{\omega_f}{\omega_f^*}\right)^2 \quad (2)$$

From Eq. (2) and ignoring the change in Re, the effective Q-factor of tip 1 is predicted as 343.1 (vs. 335.6 experimental) and tip 2 as 233.7 (vs. 244 experimental). Hence, the change in the Q-factor is primarily due to the change in the spring constant and resonance frequency due to the increase in beam thickness, while the increase in the internal damping in the compound beam is smaller than the extrinsic ambient damping.

To characterize the tip geometry, the tip was used to image a calibration grating (TGT01, Mikromasch) containing an array of sharp tips. The blind reconstruction of the tip geometry was performed using the SPIP 'tip characterization' option (Image Metrology). Tip 1 (repaired [Fig. 1(i)]) was found to have a 104 nm radius (9.5 nm x-radius and 8.7 nm y-radius) and cone angles of 35º and 42º along the x- and y-axis, respectively. The resulting geometry, including the asperity and the surrounding oxide are clearly visible in Fig. 2 (d).

The electrical properties of the conductive tip were probed using I-V measurements in contact with a gold surface (gold-coated mica substrates, Agilent Technologies). The resulting I-V curve [Fig. 2(b)] shows semiconductor-like characteristics and strong asymmetry that can be ascribed both to the semiconducting nature of the tungsten plug and the Si-W contact. On the reverse bias, the current was found to follow a Schottky-like dependence, $\ln I = a + bV$, where $a$ = -10.149 ± 0.001 and $b$ =-0.245± 3 $10^{-4}$ and I is in [nA]. The I-V curve analysis does not allow a description by simple semiconductor models, indicative of the complex character of transport in heavily doped nanoscale contacts.

To test the suitability of the probes for electrical measurements, they were utilized to measure the local electromechanical response of ferroelectric samples, a method which requires good electrical contact between the tip and sample. Topography, PFM amplitude, and PFM phase images of PPLN obtained with tip 1 are shown in Fig. 3(a,b,c), respectively. As expected, the PFM phase provides strong polarization-dependent contrast. The PFM amplitude images show non-zero response from both domains. Similar images are obtained in liquid [Fig. 3 (d-f)] While the asymmetry in the magnitude of the response is indicative of some electrostatic component to the signal, the electromechanical contrast is clear, suggesting that there is good electrical contact between the tip and the sample. Note that after measuring force-distance curves and several contact mode scans, the tips became mechanically worn as shown in Figs. 1 (g,h). Note that a broken tip [Fig. 1 (g)] can be repaired by using FIB to re-open the via, followed by another EBID fill [Fig. 1 (i)].

To summarize, we have developed a process for the fabrication of insulated probes for electrical and electromechanical imaging in liquids and determined the effect of the coating on the dynamic properties of the cantilever. The probe geometry, conductivity, and applicability for electromechanical imaging are illustrated. The EBID process allows easy variation of the probe geometry and composition, allowing for tunable properties. The use of shielded probes may allow precise control over the application and measurement of local fields in solution.

Research sponsored by the Center for Nanophase Materials Sciences (BJR, KS, SVK, and APB), Oak Ridge National Laboratory, managed and operated by UT-Battelle, LLC for the Office of Basic Energy Sciences, U.S. Department of Energy. Research was also supported through CNMS user proposal (SJ, PDR, and SVK).

**Table 1**

**Cantilever properties before and after deposition**

| Tip | Sum (V) | Resonance (kHz) | Q | k (N/m) | InvOLS (nm/V) |
|---|---|---|---|---|---|
| Tip 1 before | 4.14 | 97.76 | 288 | 9.82 | 95.43 |
| Tip 1 after | 4.25 | 103.10 | 335.6 | 13.42 | 87.78 |
| Tip 2 before | 5.61 | 66.74 | 156 | 1.78 | 75.75 |
| Tip 2 after | 5.49 | 76.14 | 244.2 | 3.59 | 79.72 |

**Figure Captions**

**Fig. 1.** SEM images (a,d) after the oxide deposition, (b,e) after the via FIB, and (c,f) after the EBID fill for tips 1 and 2, respectively. (g,h) SEM images of tips 1 and 2, respectively, after using the tips for contact mode imaging. (i) SEM of tip 1 after repairing the plug in (g). Circles in (g,h) show the damaged plugs.

**Fig. 2.** (a) First thermal resonance before and after the tip processing for tip 1, (b) I-V curve for tip 1 (repaired), (c) scan of calibration grating using tip 1 (repaired) with inset of tip reconstruction (z-scale 162.5 nm), and (d) 3D representation of tip-shape. The z-scale for (c) is 200 nm.

**Fig. 3.** (a) Topography, (b) PFM amplitude, and (c) PFM phase images of PPLN measured with tip 1. The vertical scales are 5 nm for (a) and 10 V for (c). (d,e,f) Topography, PFM amplitude, and PFM phase images of PPLN measured with tip 1 in deionized water. The vertical scales are 6nm for (d) and 5 V for (f).

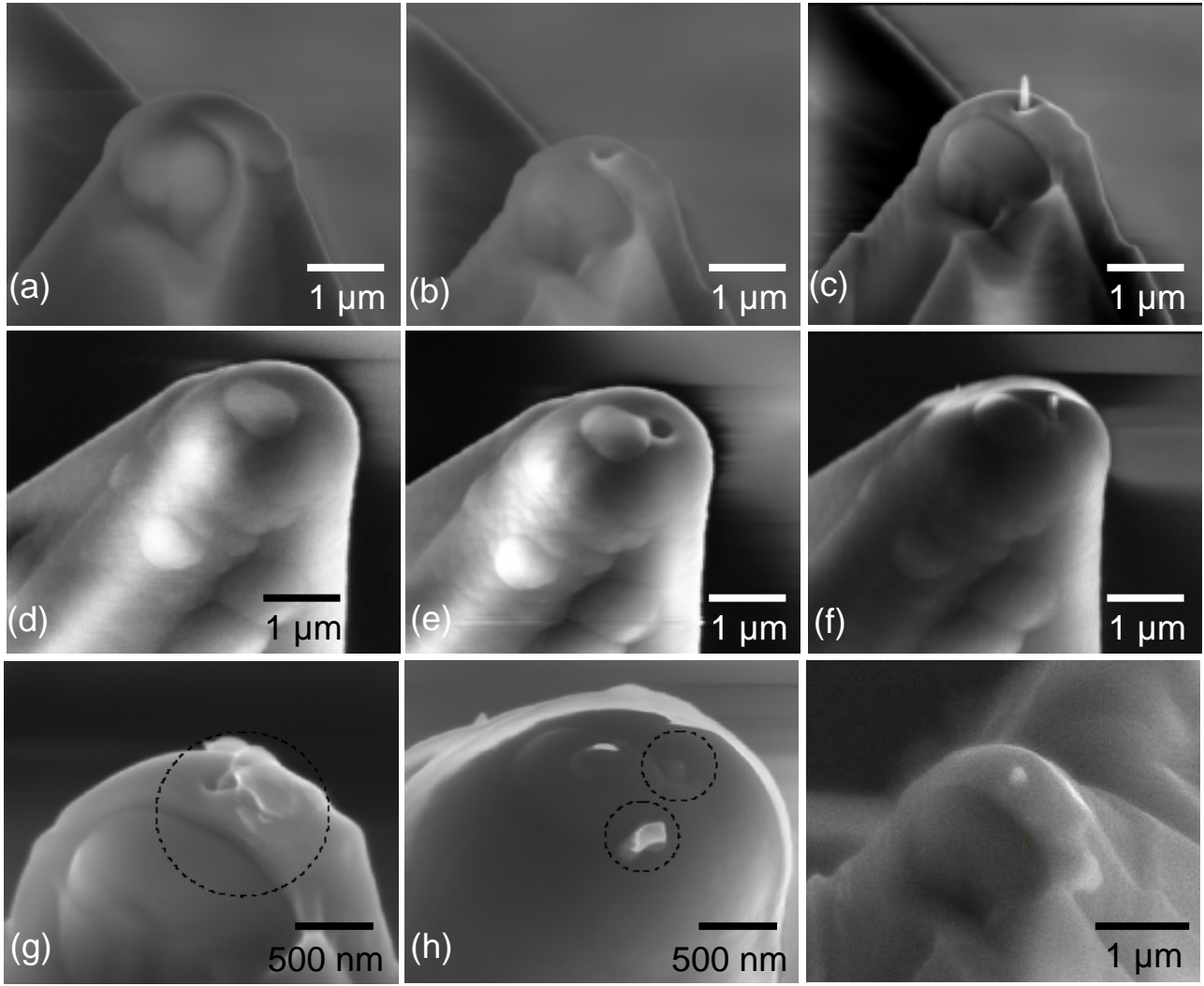

**Figure 1.** Brian J. Rodriguez et al.

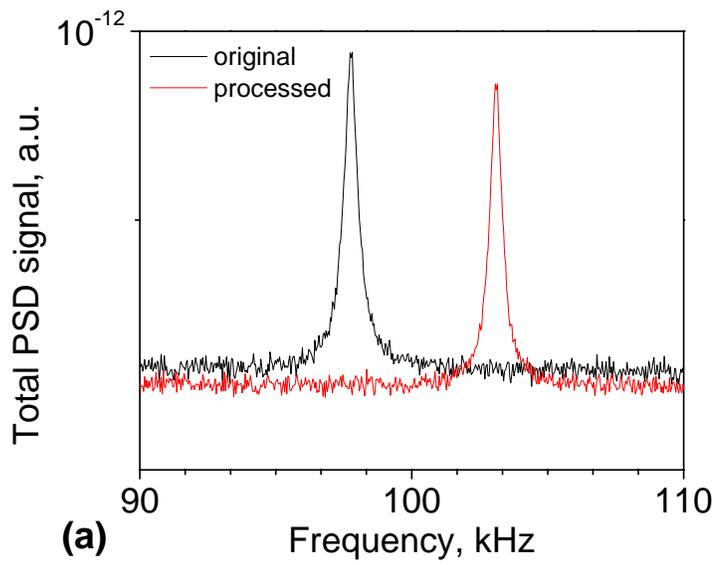
(a)

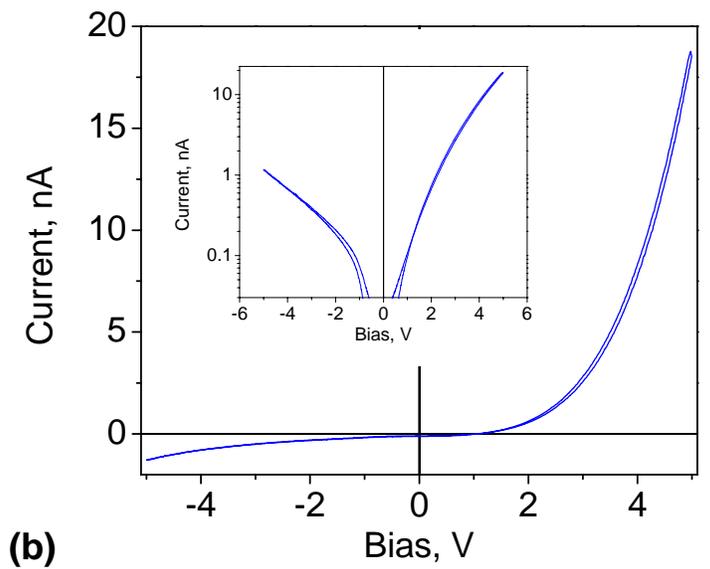
(b)

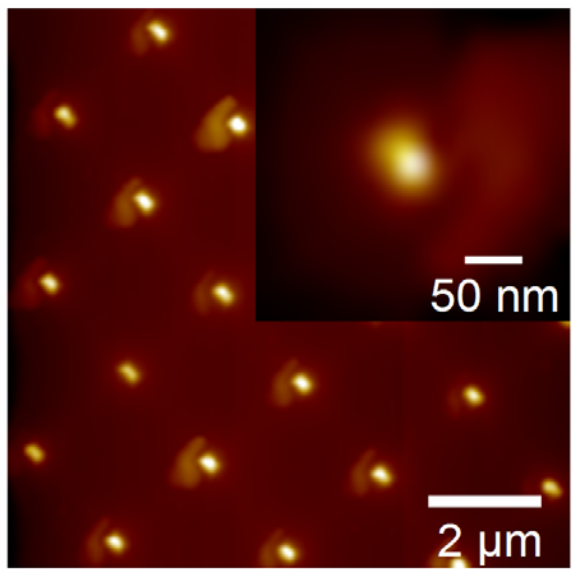
(c)

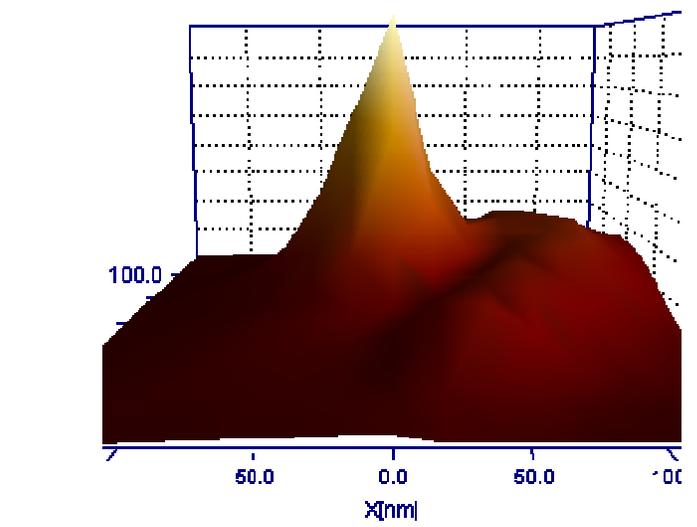
(d)

**Figure 2.** Brian J. Rodriguez et al.

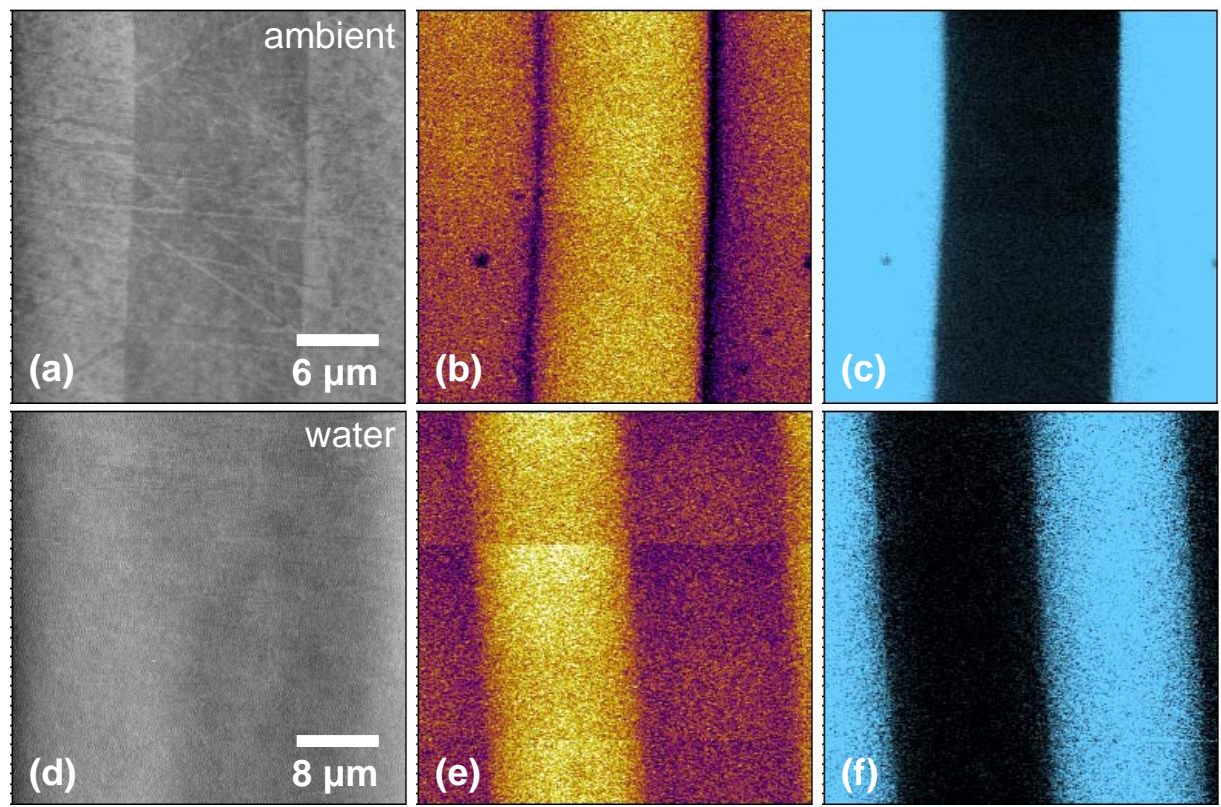

**Figure 3.** Brian J. Rodriguez et al.